\documentstyle[12pt]{article}
 
\newcommand{\be}{\begin{equation}}
\newcommand{\ee}{\end{equation}}
\def\n{\noindent}

\catcode `\@=11
\@addtoreset{equation}{section}
 
\catcode `\@=12

\title{\bf\huge Black hole : Equipartition of matter and potential energy}
\author{Naresh Dadhich\thanks{E-mail : nkd@iucaa.ernet.in} \\
{\sl Inter-University Centre for Astronomy \& Astrophysics,}\\
{\sl Post Bag 4, Ganeshkhind, Pune - 411 007, India.} 
} 
\date{}
\begin{document}
\maketitle

\begin{abstract}                      
\n {\it Black hole horizon is usually defined as the limit for existence 
of  timelike  worldline or when a spatially bound  surface  turns 
oneway  (it  is  crossable only in one direction).  It  would  be 
insightful and physically appealing to find its  characterization 
involving  an energy consideration. By employing  the  Brown-York [1]
quasilocal energy we propose a new and novel characterization  of 
the horizon of static black hole. It is the surface at which  the 
Brown-York  energy equipartitions itself between the  matter  and 
potential energy. It is also equivalent to equipartitioning of
the binding energy and the gravitational charge enclosed by the horizon.} \\ 
\end{abstract}


\vfill
\begin{flushright} IUCAA-39/97 \end{flushright}
\vfill

\newpage
\n The  intutive picture of a black hole is a compact  object  with 
very  strong  gravitational field, so strong  that  even  photons 
cannot  cross out of a spatially bound surface. That means it  is 
natural  to  expect that the  physical  arguement  characterizing 
black  hole  should be based on some kind of  energy  balance. The 
usual considerations of limit to existence of timelike  worldline 
and oneway character of a compact surface do not directly follow from  any 
energy consideration. They really appeal to the general  spacetime 
properties.  In  contrast  the  classical  arguement  of   escape 
velocity  (balance  between kinetic and potential  energy  for  a 
particle travelling with the velocity of light!) is an  excellent 
example  of  an  energy  arguement.This is  of  course  based  on 
application of a wrong formula, though giving the right result. \\
 
\n In general relativity (GR) it is not only the matter energy that 
produces  gravitational field but the gravitational field  energy 
too  does. This would however be higher order non-linear  effect. 
The   energy  of  a  system  will  therefore  consist   of   both 
contributions.  It  is a well-known fact that energy  is  a  very 
difficult  quantity  to  handle in GR,  because  its  measure  is 
inherently   ambiguous.   The  cause  for   ambiguity   is   non-
localizibility  of gravitational field energy. This is why  there 
exist  in literature several definitions for  quasilocal  energy [1-6]. 
Recently  Brown and York [1] have proposed a natural and  interesting 
definition  employing the Hamilton-Jacobi theory. It includes,  as 
one  would  like to have, both matter and potential  (which  will 
include  all  that  cannot be  included  in  the  energy-momentum 
tensor) energy. It is defined covariantly and most importantly it 
is additive, energies could be added. If we can separate out  the 
two  components, then it would be an interesting question to  ask 
what  would be characterized by their equality? It will turn  out 
to be the black hole horizon. The Brown-York energy is computable 
for  static spacetimes but not in general for the Kerr  spacetime 
as  2-surfaces  that  could  be  isometrically  embedded  in   3-
hypersurface  do  not close. We shall hence be considering the 
charged   black  hole  which  will  automatically   include   the 
Schwarzschild hole. \\
 
\n The Brown-York energy is defined by [1]

\be
E = \frac{1}{8 \pi} \int (K - K_0) \sqrt{\sigma} d^2 x
\ee

\n where $\sigma$ and $K$ are determinant of the 2-metric and mean  extrinsic 
curvature  of  the  2-surface  $B$  isometrically  embedded  in  3-
hypersurface $C$.  Here the subscript 0 refers  to  the  reference 
spacetime,  with  respect  to  which  $E$  is  being   measured. The 
quasilocal  energy $E$ is minus the variation  in the action  in  a 
unit  increase in  proper  time separation  between  $B$  and  its 
neighbouring  2-surface. Thus it is the value of the  Hamiltonian 
that generates limit time translations perpendicular to $C$ at  the 
boundary $B$. This is clearly the most natural definition and it is 
also physically  satisfactory as it includes contributions  from  both 
matter and potential energy. \\

\n For the charged black hole (1) will give 

\be
E = R - (R^2 - 2MR + Q^2)^{1/2}.
\ee

\n Here the reference spacetime is asymptotic flat spacetime. In  the 
Newtonian approximation, it will read 

\be
E = M - \frac{Q^2}{2R} + \frac{M^2}{2R}.
\ee

\n Clearly  $E$  is  the sum of matter energy  density  and  potential 
energy associated with building a charged fluid ball by  bringing 
together individual particles from some initial radius. It  could 
be  understood  in  the following way : $M$  is  the  total  energy 
including  rest  mass  and  all  kinds  of  interaction  energies 
localizable as well as nonlocalizable, the energy lying  exterior 
to  the radius $R$ will be $Q^2/2R$  arising from $T^0_0 = Q^2/2R^4$
due to electric 
field, plus the gravitational potential energy, which cannot  be 
put in $T^k_i $ without  ambiguity  and  its  first  term   in   the 
approximation  is $-M^2/2R$, hence the energy contained  inside  the 
radius  $R$ will be $M-(Q^2/2R - M^2/2R)$. This is what the 
quasilocal $E $
is.  We  have  dumped together all that which cannot  be  put  in        
$T^k_i$ and call it potential energy. \\

\n Let  us  now compute the matter energy arising from  the  energy 
density $Q^2/2R^4$  due  to electric field. Integrating it  over  from           
$\infty $ to  $R$, we get $- Q^2/2R$  add $M$ to it to write $M - Q^2/2R$, 
the  matter 
energy  component  of $E$. Equipartitioning of $E$  into  matter  and 
potential energy ($E$ -- (matter energy)), will give 

\be
R - (R^2 - 2MR + Q^2)^{1/2} = 2 (M - Q^2/2R)
\ee

\n which will imply 

\be
(2MR - Q^2) (R^2 - 2MR + Q^2) = 0.
\ee

\n This means either $R^2 = 2MR - Q^2= 0$  
which defines the black hole horizon                          
for $M^2 \geq Q^2$, or $R = Q^2/2M$,  
the  hard   core   radius for naked         
singularity  for $M^2 < Q^2 $. This is how is characterized the  horizon 
of a static (charged or otherwise) black hole marking the balance 
between  matter  and  potential energy. Note that  in  the  naked 
singularity  case, the two (matter and potential energy)  can  be 
equal only when each vanishes. \\
 
\n The  energy  of the hole at infinity will always be $ M$,
because contribution  of  electric field to matter  energy  as  well  as 
potential  energy fall off to zero asymptotically. As  one  comes 
closer  to  the hole potential energy goes  on  increasing  while 
matter  part  is  non-increasing and the two turn  equal  at  the 
horizon.  The  potential energy is created by matter  energy  and 
hence  it  can only produce second order  nonlinear  relativistic 
effect.  The  equality  of  the  two  signals  the   relativistic 
nonlinear  effect being as strong as the Newtonian  effect.  This 
will  indicate strong field limit. How could that be typified?  A 
typical  measure  of the limiting strength would be  that  gravitational
pull  becomes 
irresistable and/or nothing can tunnel out of it. This is  precisely 
the characterization of horizon. \\ 

\n It  is  the most remarkable and novel feature that  the  horizon 
equipartitions  quasilocal energy of the hole between matter  and 
potential  energy. It reminds of the classical   escape  velocity 
consideration  marking the balance between kinetic and  potential 
energy.  Here also the absolute value of gravitational  potential                            $\phi = -(M - Q^2/2R)/R$
due  to matter energy attains the value 1/2 at the horizon.  This 
is  equivalent to $E = 2(M - Q^2/2R)$ because $E$ goes as $R $ close  to 
the horizon. Intutively one can say that it is the matter  energy 
that  gives  rise to gravitational  attraction   while  potential 
energy  acts  passively  through  curving  space,  which  is  the 
relativistic   nonlinear  effect [7].  (It  is  known   that   photon 
propogation  will be affected by this nonlinear  interaction  in 
equal  measure.) When the two become equal, they join in   unison 
to  keep evrything confined to a compact 2-surface which  defines 
the horizon. What it means is that the two contribute equally 
to gravitational field at the horizon. \\  

\n It   is   also  possible  to  relate  quasilocal   energy   with 
gravitational  charge of a black hole. The charge is  defined  by 
the  flux of red-shifetd proper acceleration across a  closed  2-
surface [8-9] and  it  is in general  different  from  the  quasilocal 
energy.  It turns out that the charge enclosed in the horizon  is 
equal to the binding energy $(E(R_+) - E(\infty))$ at the horizon [10].
As a matter of fact we obtain the same characterizing eqn.(5) when
the binding energy $(E(R) - M)$ is equated to the gravitational
charge $M - Q^2/R$ [8-9].   This 
is  how charge and energy are related. The quasilocal  energy  is 
conserved,  $E = M$ everywhere, for the extremal  hole, $M^2 = Q^2$ while 
charge is conserved for the Schwazschild hole, $Q = 0$ [2]. \\
 
\n Equipartitioning  of  energy contained inside the  horizon  into 
matter  and  potential energy should be a general  principle  for 
characterization  of  black  hole and should  hence  be  true  in 
general.  Unfortunately it cannot be applied to a rotating  black 
hole straightway because it is not possible to compute $E$ for  the 
Kerr  metric in which 2-surface does not close in general. It  is 
however possible to compute gravitational charge enclosed by  the 
horizon [8-9]  and  it is given by $(M^2 - a^2)^{1/2}$. 
It may be  noted  that  the 
above  prescription  in terms of binding energy and  charge  does 
yield  the right result $E(R_+) = R_+ = M + (M^2 - a^2)^{1/2}$.
This indicates that  though  we 
cannot  compute  $E$  at any $R$, but the Kerr horizon  seems  to  be 
similarly characterized. \\

\n This is undoubtedly a very interesting and novel application  of 
the Brown-York energy. \\

\n It is a pleasure to thank Sukanta Bose for the enlightening  and 
fruitful discussions.

\end{document}